\documentstyle[psfig]{article}
\begin{document}

\begin{center}
{\large \bf Synthesis of radioactive nuclei and gamma-line radiation 
from novae}
\end{center}

\begin{center}
\bf A.D.Kudryashov$^1$, N.N.Chugai$^2$, and A.V.Tutukov$^2$\\
$^1$VINITI, Moscow, Russia\\
$^2$Institute of Astronomy of the RAS, Moscow, Russia
\end{center}

\begin{abstract} We carried out kinetic calculations of thermonuclear
burning in the hydrogen-rich matter to simulate nucleosynthesis yields
in nova outbursts. These results are used to calculate the light curves
of annihilation gamma-ray line from N, O and F radioactive isotopes.
\end{abstract}

Thermonuclear runaway on accreting white dwarfs (WD) is a conventional
model of nova outbursts. High temperature hydrogen burning
converts most of initial CNO nuclei to radioactive isotopes $^{13}$N,
$^{14}$O, $^{15}$O, $^{17}$F and $^{18}$F (NOF isotopes); their
presence in novae may be evidenced by 511 keV annihilation emission 
(Clayton \& Hoyle 1974; Leising \& Clayton 1987). 
Radioactive $^{22}$Na also synthesized
in novae may be detected in 1.275 MeV line. Furthermore,
galactic $^{26}$Al already observed
in 1.809 MeV line may be also contributed by novae. 
Here we concentrate only
on the synthesis of short-lived NOF isotopes and computation of
annihilation line flux from novae.

Following Kudryashov \& Tutukov (1995) we calculate nucleo\-syn\-the\-sis 
in novae using one-zone model.
Temperature and density are assumed constant during the burning
which is terminated with an exhaustion of the hydrogen fraction 
$\Delta$X=0.1. The temperature is taken from the range $(1-3)\cdot10^8$ K, 
while density is $10^4$ g cm$^{-3}$ in all cases. 
The adopted composition of CO dwarf is 
X($^{12}$C) = 0.49, X($^{16}$O) = 0.49, X($^{22}$Ne) = 0.01,
X($^{25}$Mg) = 0.01, while for ONeMg dwarf the composition is
X($^{16}$O) = 0.3, X($^{20}$Ne) = 0.5, X($^{24}$Mg) = 0.2.
Mixing parameter $q$ (mass fraction of
the WD matter in the total envelope mass) is varied in the range
0.1--0.9. A kinetic network adopted in this paper is an updated 
version of that
from Kudryashov \& Tutukov (1995). Apart from H and He, it includes nuclei 
from C to Ca and all nuclear reactions with charged particles. 

The obtained amount of NOF isotopes in both types of WD 
is roughly equal to total mass of
admixed CO matter of WD in the envelope and thus is proportional to
the mixing parameter $q$ (Fig. 1). 
The fraction of $^{18}$F in most cases is within the range 
$10^{-3}-10^{-2}$.
The mass fraction of $^{22}$Na in ONeMg WD envelopes exceeds 1\%
in a certain region of $T-q$ plane, 
while in CO dwarf envelopes the fraction of $^{22}$Na
is usually lower than $10^{-3}$. The fraction of $^{26}$Al 
on ONeMg WD for reasonable values of $T$ and $q$ 
is between $10^{-4}-10^{-2}$, while on
CO WD $^{26}$Al content is in the range $10^{-5}-10^{-3}$.

We summarize some results of nucleosynthesis computations in Table 1,
where compositions of major radioactive isotopes are given for 
average values of $T$ and $q$ parameters (typical case) and for 
parameters, which favour the maximum production of $^{18}$F 
(optimistic case). We suggest that computed isotope composition 
refers to the burning zone, which presumably occupies
a fraction $\psi$ of the envelope mass. This fraction is
computed from the energy balance (nuclear energy
is the sum of the gravitation binding and kinetic 
energy) for 1$M_{\odot}$ WD. Prior to the ejection, 
isotopes are presumably mixed in the inner fraction 
$f_{\rm mix}$ of the envelope. 

\begin{table}
\caption{Isotope composition for typical ($T_8=2$, $q=0.5$, top) 
and optimistic ($T_8=1$, $q=0.7$, bottom) cases}
\vspace*{1mm}
\footnotesize
\tabcolsep=1.3mm
\begin{tabular}{ccccccccc}
\hline
WD  & $\epsilon$, $10^{17}$ &  $^{13}$N & $^{14}$O & $^{15}$O & $^{17}$F & $^{18}$F &
$^{22}$Na & $^{26}$Al \\
  &  erg g$^{-1}$ &    (862 s)  & (102 s) & (176 s) & (95 s) & (158 min) & (3.75 yr) &
($1.04\cdot10^6$ yr) \\
\hline
CO     & 4.5 & 4.8e-4 & 2.1e-1 & 2.5e-1  & 8.4e-2 & 2.4e-3 & 2.3e-4 & 9.4e-5\\
ONeMg  & 5.6  & 1.6e-4 & 1.1e-2 & 8.1e-2 & 6.0e-2 & 1.7e-3 & 1.6e-2 & 2.0e-3\\
\hline
CO     & 6.1 & 1.0e-1 & 2.1e-2 & 5.8e-2  & 5.4e-3 & 7.4e-3 & 2.5e-7 & 5.5e-6\\
ONeMg  & 6 & 2.3e-2 & 4.2e-3 & 1.4e-2 & 1.0e-3 & 3.9e-2 & 2.0e-4 & 6.1e-4\\
\hline
\end{tabular}
\end{table}

The flux in the annihilation line for adopted abundances of 
radioactive isotopes in the typical case (Table 1) is 
computed assuming that ejecta may expand either
homologously ($v=r/t$) or in the form of
wind outflow with constant velocity and mass-loss rate.
The ejecta mass, velocity and $f_{\rm mix}$ are free parameters. 
In the wind case the outflow kinetic luminosity is fixed at Eddington
limit $L_{\rm k}=(1/2)\dot{M}v^2=10^{38}$ erg s$^{-1}$.
The emergent annihilation luminosity is determined by the mass of
a transparent outer layer ($\tau\leq 1$). The density dependence 
of a probability of the two-photon positronium
annihilation is taken into account.

Given equal outer velocity of ejecta, the emergent luminosity of
gamma-rays is determined by the density distribution of outer layers. 
The homologous homogeneous sphere 
(model HH1, see Table~2, Fig.~2a) produces slightly 
higher second maximum ($^{18}$F) compared to the 
homologous envelope with the power law density
distribution $\rho \propto v^{-7}$ (model HP).
This is caused by the lower transparent mass in HP model. 
The wind model W gives considerably lower
flux due to the lower density.

\begin{table}
\caption{Parameters of novae and annihilation line flux ($d=1$ kpc)}
\vspace*{1mm}
\footnotesize
\tabcolsep=2mm
\begin{tabular}{ccccccccc}
\hline
Model & WD & $M$ & $V_{\rm max}$  & $f_{\rm mix}$ & $\psi$  & $F$ &
$\Phi$ & $\Delta t$\\
  &  &  $10^{-5}M_{\odot}$ & km s$^{-1}$ &   &
& $10^{-4}$ cm$^{-2}$ s$^{-1}$ & cm$^{-2}$ & $10^4$ s \\
\hline
HP   & CO &  2  & 2500    & 1    & 0.31 & 19  & 53  & 2.8 \\
HH1 &  CO &  2  & 2500    & 1    & 0.34 & 33  & 107 & 3.2 \\
W   &  CO &  2  & 2500    & 1    & 0.28 & 1   & 2.6 & 2.6 \\
HH2 &  CO &  2  & 2500    & 0.34 & 0.34 & 0.3  & 1  & 3.7 \\
HH3 &  CO &  2  & 2500    & 0.99 & 0.34 & 28  & 85  & 3.1 \\
HH4 &  CO &  10 & 2500    & 1    & 0.34 & 34  & 116 & 3.4 \\
HH5 &  CO &  2  & 3500    & 1    & 0.38 & 70  & 214 & 3 \\
HH6 & ONeMg &  2  & 2500  & 1    & 0.25 & 17  & 56  & 3.2\\
HH7 &  CO &  2  & 2500    & 1    & 0.34 & 83  & 280 & 3.4 \\
HH8 & ONeMg &  2  & 2500  & 1    & 0.25 & 400 & 1300 & 3.3\\
\hline
\end{tabular}
\end{table}

The effect of variation of mass, velocity, composition (ONeMg WD
vs. CO WD), and degree of mixing is displayed
at Fig. 2b (HH1--HH6 models). The minimum mixing degree
restricted by the burning zone ($f_{\rm mix}=\psi=0.34$) 
severely suppresses both light curve maxima. Remarkably, 
the marginally incomplete mixing, 
$f_{\rm mix}=0.99$, with only 1\% of unmixed outer shell 
results in the complete suppression of initial part ($t<10^4$ s) 
of light curve. Therefore, the 
incomplete mixing, which is quite conceivable in realistic novae
makes the detection of the first 
maximum ($t<10^3$ s) very problematic.

The Table 2 shows also flux $F$ (distance 1 kpc) 
in the second maximum ($t\approx 2\cdot10^4$ s) related to 
$^{18}$F, integrated flux (fluence $\Phi$) and the characteristic
width ($\Delta t=\Phi/F$). Summing up these numbers, 
for typical burning case 
the expected fluxes from novae at 1 kpc are in the range  
$3\cdot10^{-5} - 7\cdot10^{-3}$ cm$^{-2}$ c$^{-1}$. The upper
limit exceeds by a factor of two the detection limit of 
BATSE for 0.5 day-long events (Smith et al. 1996).
Given the fact that roughly every three year nova may
occur at the distance $d<1$ kpc the detection of
annihilation line from novae seems quite plausible.
The flux from the optimistic 
ONeMg case (HH8, Table 2), may be as high as
$4\cdot10^{-2}$ cm$^{-2}$ c$^{-1}$ making such events
mostly favorable for detection. 

\section*{\bf References}

\parindent 0pt

Clayton D.D., Hoyle F., 1974, ApJ, 187, L101

Leising M.D., Clayton D.D., 1987, ApJ, 323, 159

Kudryashov A.D., Tutukov A.V., 1995, Astron. Reports, 39, 482

Smith D.M., Leventhal M., Cavallo R., et al., 1996, ApJ, 471, 783

\begin{center}
\begin{figure}
\psfig{file=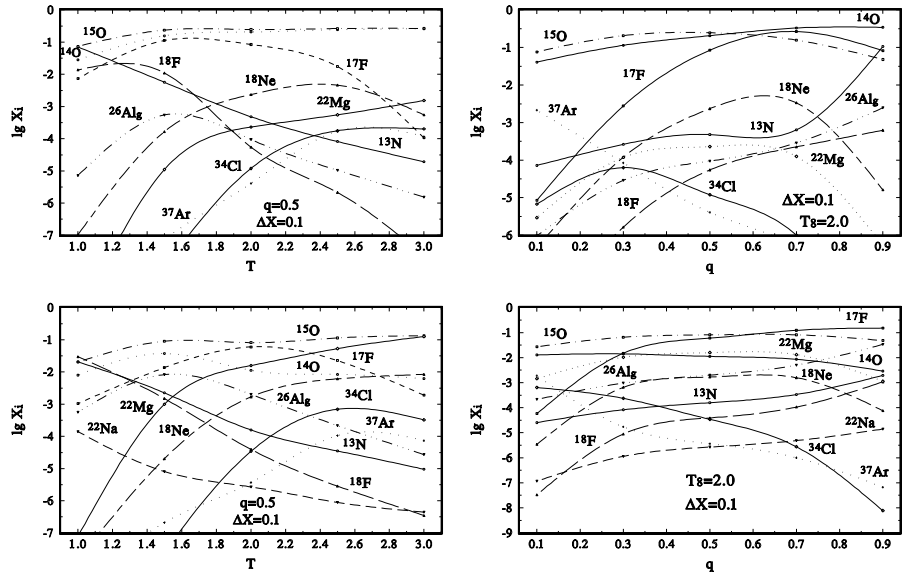,width=14.5cm}
\caption{ Radioactive isotope mass fractions in the envelopes 
of CO WD (top) and ONeMg WD (bottom).
On the left panel is the temperature dependence for $q=0.5$, while shown on
the right panel is the mixing parameter dependence for $T_8=2$.}
\end{figure}
\end{center}

\begin{figure}
\begin{center}
\psfig{file=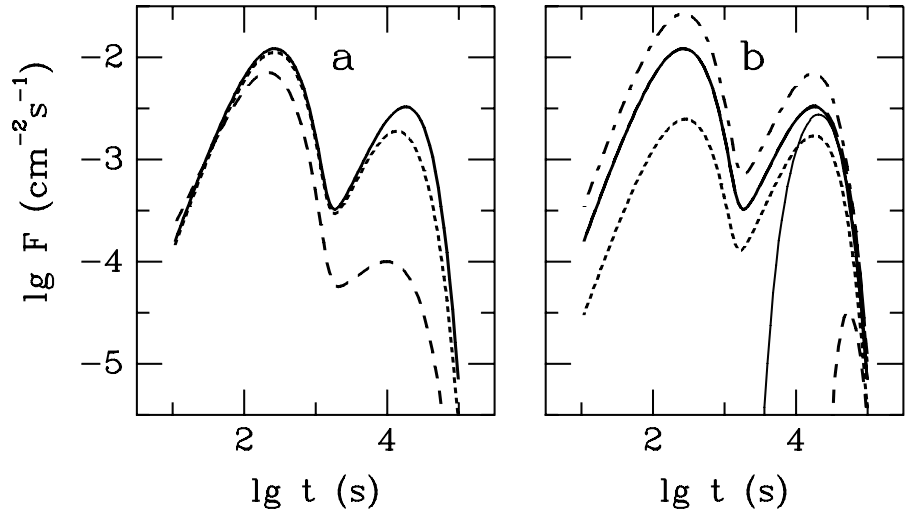,height=6.7cm}
\end{center}
\caption{ Annihilation line flux from model novae. The left 
panel (a) shows the effect of different kinematics and density distribution
for homologous models HH1 (solid line) and HP (dotted), as well
as for the wind outflow model W (dashed).
The right panel (b) shows the effect of the variation of parameters
and composition compared to the template model HH1 (thick solid line), viz.
model with no mixing (HH2, short dashes), marginally homogeneous mixing 
(HH3, thin solid line), higher ejecta mass (HH4, long dashes), higher
velocity (HH5, dash-dotted), ONeMg WD (vs. CO WD) composition (HH6, dotted). 
}
\end{figure}

\end{document}